\documentclass[12pt]{article}
\usepackage{amsmath}
\usepackage{graphicx}
\begin{document}
\baselineskip=18 pt
\begin{center}
{\large{\bf Weyl Fermions in a Linear Class of G\"{o}del-type Space-Time Backgrounds}}
\end{center}

\vspace{.5cm}

\begin{center}
{\bf Faizuddin Ahmed}\footnote{\bf faizuddinahmed15@gmail.com ; faiz4U.enter@rediffmail.com}\\ 
{\bf National Academy Gauripur, Assam, 783331, India}
\end{center}

\vspace{.5cm}

\begin{abstract}

In this paper, we study Weyl fermions in a linear class of topologically trivial G\"{o}del-type geometries in the Einstein's general relativity. We solve the Weyl equation and evaluate the energy eigenvalues and corresponding wave-functions, in detail.

\end{abstract}

{\bf keywords:} Exact solution of field equations, G\"{o}del-type space-time, Relativistic wave-equations, energy spectrum, eigenfunctions, vorticity.

\vspace{0.1cm}


\section{Introduction}

The particle motions are commonly described using either by the Klein-Gordon or the Dirac equation \cite{XCZ,LZY} depending on spin character of the particle. The spin-zero particles like mesons, bosons are described by the Klein-Gordon equation, and spin-half particles such as electron is described satisfactorily by the Dirac equation. The Dirac and Klein-Gordon equations have been of interest for theoretical physicists in many branches of physics \cite{AWT,BT}. Since the exact solutions of the Klein–Gordon and the Dirac equation play an important role in the relativistic quantum Physics as well as in various physical applications including those in nuclear and high energy physics \cite{TYW,WG}.

Several studies of the physical problems involving G\"{o}del-type space-times have been developed in recent years. Recently, a large number of issues related to rotating G\"{o}del solutions in the general relativity theory as well as alternative theories of gravitation have been studied ({\it e. g.}, \cite{JS,CF,CF2,CF3,JB3,JA,Caval,GC,GC2,AHEP}). The relativistic quantum dynamics of spin-$0$ and spin-half particles have been investigated in G\"{o}del-type space-times in the relativistic quantum mechanics by many authors. For examples, Figueiredo {\it et al.} \cite{BDB} had investigated the scalar and spin-$1/2$ particles in a G\"{o}del space-times with positive, negative and zero curvatures. The relationship between the Klein-Gordon solution in a class of G\"{o}del solutions in the general relativity theory with the Landau levels in a curved spaces were investigated in Refs. \cite{ND,ND2}. This analogy was also observed by Das {\it et al.} \cite{SD} studying the quantum dynamics of a scalar particle in the Som-Raychaudhuri space-time, and compared with the Landau levels in flat space. In Ref. \cite{Wang}, the relativistic quantum dynamics of a scalar quantum particle in the Som-Raychaudhuri space-time were investigated. In Ref. \cite{JC}, a scalar quantum particle in a G\"{o}del-type metrics with cosmic string passing through axis have been investigated. In Ref. \cite{VM}, solution of the Weyl equation in a non-stationary G\"{o}del-type cosmological universe were obtained, and also the Weyl equation in a family of G\"{o}del-type metrics were studied. In Ref. \cite{SGF}, the Klein-Gordon equation for a particle confined in two concentric thin shells in G\"{o}del, Kerr-Newman and FRW space-times in the presence of topological defects were studied. In Ref. \cite{AH}, solution of photon equation in a stationary G\"{o}del and G\"{o}del-type space-times were obtained.

In this article, we investigate the relativistic quantum dynamics of fermion without topological defects in a class of flat or linear G\"{o}del-type metric. The purpose of this contribution is to investigate the influence of curvature, and the rotation of G\"{o}del-type metrics in the quantum dynamics of Weyl fermions. In Ref. \cite{GQ}, quantum dynamics of Dirac fermions in a G\"{o}del-type geometry of positive, negative and flat curvature with topological defects and torsion were investigated. In Ref. \cite{GQG}, the relativistic quantum dynamics of a massless fermion in the presence of topological defects in a class of G\"{o}del-type metrics were investigated. They solved the Weyl equation in the Som-Raychaudhury, spherical and hyperbolic background metrics pierced by topological defects. In Ref. \cite{EPJC}, the relativistic quantum dynamics of a spin-$0$ scalar particle in a linear class of topologically trivial G\"{o}del-type space-time was investigated. In Ref. \cite{EPJC2}, linear and Coulomb confinement of a scalar particle in a linear class of topologically trivial G\"{o}del-type space-time was investigated. In Ref. \cite{EPJC3}, the Dirac equation in a linear class of topologically trivial G\"{o}del-type space-time backgrounds was investigated. In Ref. \cite{MPLA}, spin-$0$ system of the DKP equation in a linear class of topologically trivial G\"{o}del-type space-time backgrounds was investigated. In Ref. \cite{CTP}, the relativistic quantum dynamics of spin-$0$ system of the DKP oscillator in a linear class of G\"{o}del-type space-time was studied. The results obtained for the present case of Weyl fermion in a linear class of G\"{o}del-type metrics in the Einstein's relativity theory are quite different from the previous result obtained for Dirac fermion with torsion in Ref. \cite{GQ}, and for Weyl fermions with topological defect in Ref. \cite{GQG}. The generalized KG-oscillator with position dependent mass in a linear or flat class of G\"{o}del-type space-time were studied in Ref. \cite{Y}. In the present system, the model we present is described by the Weyl fermions where Fermi velocity play the role of the speed of light in this effective theory. The study carried out in this article can be used to investigate the influence of rotation, and curvature in the condensed matter systems described by massless fermion. The approach applied in this paper can be used to investigate the Hall effect as studied  in spherical droplets \cite{BI} with rotation in the absence of topological defects. We claim that the studies of this problem in the present paper can be used to investigate the influence of rotation in the G\"{o}del-type background metrics. In Ref. \cite{Caval}, influence of rotation in fullerene molecule was investigated where in this model the rotation is introduced via a three-dimensional Godel-type metric. In this paper, we analyse the relativistic quantum dynamics of a massless fermion in a linear class of G\"{o}del-type metrics. We find the eigenvalues of energy and observe their similarity with the Landau levels for a massless spin-1/2 particle on the same space-time geometries. The possibility of zero mode for the eigenvalues of the Weyl spinor are discussed and the physical implications are analysed in this paper.

We will introduce the Dirac equation in Weyl representation on a curved background. Following the theory of spinors in curved space-time \cite{Naka,Weinberg,Cartan}, one must can write the equations for massless spin-half field in the follow way,
\begin{eqnarray}
\label{1}
&&i\,\gamma^{\mu} ({\bf x})\,\nabla_{\mu}\,\Psi=0\quad (\hbar=1=c),\\
\label{2}
&& (1+\gamma^{5})\,\Psi=0,
\end{eqnarray}
here we have that $\gamma^{\mu} ({\bf x})=e^{\mu}_{(a)}\,\gamma^{(a)}$ are the gamma matrices in Weyl representation and $\gamma^{5}=i\,\gamma^{0}\,\gamma^{1}\,\gamma^{2}\,\gamma^{3}$.

The Dirac matrices will be
\begin{equation}
\gamma^{0}=\left (\begin{array}{ll} 
{\bf I} & \,\,\,{\bf 0} \\
{\bf 0} &  -{\bf I}
\end{array} \right),\quad
\gamma^{i}=\left (\begin{array}{llll}
\,\,{\bf 0} & \sigma^i \\
-\sigma^i & {\bf 0}
\end{array} \right),
\label{3}
\end{equation}
where $I$ is a $2\times 2$ unit matrix, ${\bf 0}$ is the null matrix and $\sigma^i$ are the Pauli matrices given by
\begin{equation}
I=\left (\begin{array}{ll}
1 & 0\\
0 & 1
\end{array} \right),\quad 
\sigma^{1}=\left (\begin{array}{ll}
0 & 1\\
1 & 0
\end{array} \right),\quad 
\sigma^{2}=\left (\begin{array}{ll}
0 & -i\\
i & \,0
\end{array} \right),
\sigma^{3}=\left (\begin{array}{ll}
1 & \,\,0\\
0 & -1
\end{array} \right).
\label{4}
\end{equation}

\section{Weyl fermions in a linear class of G\"{o}del-type space-time}

Consider the following stationary space-time \cite{Faiz} (see also, Refs. \cite{AHEP,EPJC,EPJC2,EPJC3,MPLA,CTP,Y}) in the coordinates $(x^0=t, x^1=x, x^2=y, x^3=z)$ given by
\begin{equation}
ds^2=-dt^2+dx^2+\left(1-\alpha_{o}^2\,x^2\right)dy^2-2\,\alpha_{0}\,x\,dt\,dy+dz^2,
\label{6}
\end{equation}
where $\alpha_0 > 0$ is a real number. The ranges of the coordinates are
\begin{equation}
-\infty < t < \infty,\quad -\infty < x < \infty,\quad -\infty < y < \infty,\quad -\infty< z < \infty.
\label{7}
\end{equation}
The determinat of the metric tensor $det g=-1$ and the metric tensor for the space-time (\ref{6}) to be 
\begin{equation}
g_{\mu\nu} ({\bf x})=\left (\begin{array}{llll}
-1 & 0 & -\alpha_0\,x & 0 \\
\quad 0 & 1 & \quad 0 & 0 \\
-\alpha_0\,x & 0 & (1-\alpha_{0}^2\,x^2) & 0 \\
\quad 0 & 0 & \quad 0 & 1
\end{array} \right)
\label{8}
\end{equation}
with its inverse 
\begin{equation}
g^{\mu\nu} ({\bf x})=\left (\begin{array}{llll}
-(1-\alpha_{0}^2\,x^2) & 0 & -\alpha_0\,x & 0 \\
\quad\quad 0 & 1 & \quad 0 & 0 \\
\quad-\alpha_0\,x & 0 & \quad 1 & 0 \\
\quad\quad 0 & 0 & \quad 0 & 1
\end{array} \right)
\label{9}
\end{equation}
Using the definition of $e^{\mu}_{(a)}$ and $e^{(a)}_{\mu}$, we have
\begin{equation}
e^{(a)}_{\mu} ({\bf x})=\left (\begin{array}{llll}
1 & 0 & \alpha_0\,x & 0 \\
0 & 1 & 0 & 0 \\
0 & 0 & 1 & 0 \\
0 & 0 & 0 & 1
\end{array} \right),\nonumber
\label{10}
\end{equation}
\begin{equation}
e^{\mu}_{(a)} ({\bf x})=\left (\begin{array}{llll}
1 & 0 & -\alpha_0\,x & 0 \\
0 & 1 & \quad 0 & 0 \\
0 & 0 & \quad 1 & 0 \\
0 & 0 & \quad 0 & 1
\end{array} \right).
\label{11}
\end{equation}
which must satisfy
\begin{equation}
g_{\mu\nu} (x)= e^{(a)}_{\mu} (x)\,e^{(b)}_{\nu} (x)\,\eta_{(a)(b)},
\label{12}
\end{equation}
where $\eta_{(a)(b)}=\mbox{diag} (-1,1,1,1)$ is the Minkowski flat space metric. 

The spin connections can be determined using Christoffel symbols $\Gamma^{\mu}_{\nu\sigma}$ given in Ref. \cite{EPJC3} with the definition
\begin{equation}
\omega_{\mu\,(a)(b)} ({\bf x})=\eta_{(a)(c)}\,e^{(c)}_{\nu}\,e^{\tau}_{(b)}\,\Gamma^{\nu}_{\tau\mu}-\eta_{(a)(c)}\,e^{\nu}_{(b)}\,\partial_{\mu}\,e^{(c)}_{\nu}.
\label{17}
\end{equation}
And these are
\begin{eqnarray}
&&\omega_{t} ({\bf x})=\left (\begin{array}{llll}
0 & \quad 0 & 0 & 0 \\
0 & \quad 0 & \frac{\alpha_0}{2} & 0 \\
0 & -\frac{\alpha_0}{2} & 0 & 0 \\
0 & \quad 0 & 0 & 0
\end{array} \right),\nonumber
\label{19}
\end{eqnarray}
\begin{eqnarray}
&&\omega_{x} ({\bf x})=-\frac{3\,\alpha_0}{2}\,\left (\begin{array}{llll}
\alpha_0\,x & \quad 0 & 1 & 0 \\
0 & \quad 0 & 0 &  0 \\
-1 & \quad 0 & 0 & 0 \\
0 & \quad 0 & 0 & 0
\end{array} \right),\nonumber
\label{20}
\end{eqnarray}
\begin{eqnarray}
&&\omega_{y} ({\bf x})=\left (\begin{array}{llll}
0 & -\frac{\alpha_0}{2} & \quad 0 & 0 \\
\frac{\alpha_0}{2} & \quad 0 & \alpha_{0}^2\,x & 0 \\
0 & -\alpha_{0}^2\,x & \quad 0 & 0 \\
0 & \quad 0 & \quad 0 & 0
\end{array} \right),\nonumber
\label{21}
\end{eqnarray}
\begin{eqnarray}
&&\omega_{z} ({\bf x})=\left (\begin{array}{llll}
0 & 0 & 0 & 0 \\
0 & 0 & 0 & 0 \\
0 & 0 & 0 & 0 \\
0 & 0 & 0 & 0
\end{array} \right).
\label{22}
\end{eqnarray}
Using Eq. (\ref{17}), we can use chiral representation \cite{Pal}, so that the spinorial connection is define by
\begin{equation}
\Gamma_{\mu}=\frac{1}{8}\,\omega_{\mu\,(a)(b)} {(\bf x)}\,[\sigma^{a}, \sigma^{b}].
\label{23}
\end{equation}
And these are 
\begin{eqnarray}
\Gamma_{t} ({\bf x})&=&\frac{i\,\alpha_0}{4}\,\sigma^3,\nonumber\\
\Gamma_{x} ({\bf x})&=&{\bf 0},\nonumber\\
\Gamma_{y} ({\bf x})&=&\frac{i\,\alpha^{2}_0\,x}{4}\,\sigma^3,\nonumber\\
\Gamma_{z} ({\bf x})&=&{\bf 0}.
\label{24}
\end{eqnarray}
The generalized Pauli matrices $\sigma^{\mu} ({\bf x})=e^{\mu}_{(a)} ({\bf x})\,\sigma^{a}$ in curved space-time are
\begin{eqnarray}
\sigma^{t} ({\bf x})&=&\sigma^{0}-\alpha_0\,x\,\sigma^2,\nonumber\\
\sigma^{x} ({\bf x})&=&\sigma^1,\nonumber\\
\sigma^{y} ({\bf x})&=&\sigma^2,\nonumber\\
\sigma^{z} ({\bf x})&=&\sigma^3.
\label{25}
\end{eqnarray}

The Weyl equation will assume the following form:
\begin{equation}
i\,\sigma^{\mu} ({\bf x})\,(\partial_{\mu}+\Gamma_{\mu})\,\Psi=0.
\label{30}
\end{equation}
Having the generalized Dirac matrices and spinorial connections derived earlier, we have
\begin{equation}
\sigma^{\mu} ({\bf x})\,\Gamma_{\mu} ({\bf x})=\frac{i\,\alpha_0}{4}\,\sigma^3.
\label{31}
\end{equation}
Since the given metric (\ref{6}) is independent of $t, y, z$. Suppose the wave function to be
\begin{equation}
\Psi_0 (t,x,y,z)=e^{i\,(-E\,t+l\,y+k\,z)}\left (\begin{array}{c}
\psi_{1} (x)\\
\psi_{2} (x)
\end{array} \right).
\label{32}
\end{equation}
From Eq. (\ref{30}) we arrive at the following equations :
\begin{equation}
i\,[\sigma^0\,\partial_{t}+\sigma^2\,(\partial_{y}-\alpha_0\,x\,\partial_{t})+\sigma^1\,\partial_{x}+\sigma^3\,\partial_{z}]\,\Psi=\frac{\alpha_0}{4}\,\sigma^3\,\Psi
\label{33}
\end{equation}
Now we do a transformation
\begin{equation}
    \Psi=e^{-i\,\frac{\alpha_0}{4}\,(z-z_0)}\,\Psi_0
    \label{34}
\end{equation}
into the Eq. (\ref{33}), we get
\begin{equation}
    i\,\sigma^0\,\partial_{t}\,\Psi_0+i\,\sigma^2\,(\partial_y-\alpha_0\,x\,\partial_t)\,\Psi_0+i\,\sigma^1\,\partial_x\,\Psi_0+i\,\sigma^3\,\partial_z\,\Psi_0=0.
    \label{35}
\end{equation}
It is possible to rewrite the Weyl equation (\ref{35}) in the following way
\begin{equation}
i\,\frac{\partial\,\Psi_0}{\partial t} = \frac{1}{(1-\sigma^2\,\alpha_0\,x)}\,[\sigma^1\,\hat{\pi}_x+\sigma^2\,\hat{\pi}_y+\sigma^3\,\hat{\pi}_z]\,\Psi_0=\hat{H}\,\Psi_0,
\label{36}
\end{equation}
where $\hat {H}$ is the Hamiltonian of the Weyl particle and the conjugated momentum is
given by
\begin{eqnarray}
\hat{\pi}_x&=&-i\,\frac{\partial}{\partial x},\nonumber\\
\hat{\pi}_y&=&-i\,\frac{\partial}{\partial y},\nonumber\\
\hat{\pi}_z&=&-i\,\frac{\partial}{\partial z}.
\label{37}
\end{eqnarray}
To solve Eq. (\ref{35}), we must substitute the ansatz (\ref{32}) and the Pauli metrices (\ref{4}). Then we obtain two coupled differential equations,
\begin{eqnarray}
\label{38}
[E-k]\,\psi_1&=&-i\,[(\alpha_0\,E\,x+l)\,\psi_2+\psi'_{2}],\\
\label{39}
[E+k]\,\psi_2&=&i\,[(\alpha_0\,E\,x+l)\,\psi_1-\psi'_{1}].
\end{eqnarray}
We are capable of converting these two differential equations of first-order to two differential equations of second-order as below
\begin{equation}
    \frac{d^2\,\psi_i}{dx^2}+[\lambda_s-\beta^2\,x^2-\eta\,x]\,\psi_i=0,
    \label{40}
\end{equation}
where
\begin{eqnarray}
\lambda_s&=&E^2-\alpha_0\,E\,s-k^2-l^2,\nonumber\\
\beta&=&\alpha_0\,E,\nonumber\\
\eta&=&2\,\beta\,l,\quad s=\pm\,1.
\label{41}
\end{eqnarray}
Note that $\psi_1 (x)$ and $\psi_2 (x)$ are the wave function of $\sigma^3$ with eigenvalues $\pm 1$, so we can write $\psi_s=(\psi_{+}, \psi_{-})^T$ with $\sigma^3\,\psi_{s}=s\,\psi_{s}$, $s=\pm\,1$.

Transforming a new variable $r=\sqrt{\beta}\,x$ into the equation (\ref{40}), we get
\begin{equation}
\psi''_{i} (r) + [\frac{\lambda_s}{\beta}-r^2-\frac{\eta}{\beta^{\frac{3}{2}}}\,r]\,\psi_{i} (r)=0.
\label{42}
\end{equation}
The asymptotic behaviour of the possible solution to the Eq. (\ref{42}) are to be determined for   $r\rightarrow 0$ and $r \rightarrow \infty$. These conditions are necessary since the wave functions must be well-behaved in this limits, and thus, bound states of energy eigenvalues can be obtained. Let us impose requirement that the function $\psi_{i} (r)$ vanish at $r\rightarrow 0$ and $r \rightarrow \infty$. 
Let the solution 
 is given by
\begin{equation}
\psi_{i} (r)=r^{A}\,e^{-(B\,r+D\,r^2)}\,H (r).
\label{43}
\end{equation}
Substituting the above solution (\ref{43}) into the equation (\ref{42}), we get
 \begin{eqnarray}
   &&H'' (r)+[\frac{2\,A}{r}-2\,B-4\,D\,r]\,H'(r)+[\frac{A^2-A}{r^2}-\frac{2\,A\,B}{r}-4\,A\,D-2\,D\nonumber\\
   &+&\frac{\lambda}{\beta}+B^2+(4\,B\,D-\frac{\eta}{\beta^{\frac{3}{2}}})\,r+(4\,D^2-1)\,r^2]\,H (r)=0.
  \label{44}
 \end{eqnarray}
Equating the coefficients of $r^{-2}, r, r^{2}$ equals to zero in the above differential equation, we get
\begin{eqnarray}
 &&A^2-A=0\Rightarrow  A=1, \quad  A\neq 0,\nonumber\\
 &&4\,B\,D-\frac{\eta}{\beta^{\frac{3}{2}}}=0\Rightarrow B=\frac{1}{4\,D}\,\frac{\eta}{\beta^{\frac{3}{2}}},\nonumber\\
 &&4\,D^2-1=0\Rightarrow D=\frac{1}{2}.
 \label{45}
\end{eqnarray}
With these the above equation (\ref{44}) can be express as
\begin{equation}
H''(r)+[\frac{\gamma}{r}-\zeta-\delta\,r]\,H' (r)+[-\frac{q}{r}+\theta]\,H (r)=0,
\label{46}
\end{equation}
where
\begin{eqnarray}
&&\gamma=2\,A,\nonumber\\
&&\zeta=2\,B,\nonumber\\
&&\delta=4\,D,\nonumber\\
&&q=2\,A\,B,\nonumber\\
&&\theta=B^2+\frac{\lambda_s}{\beta}-4\,A\,D-2\,D.
\label{47}
\end{eqnarray}
The Eq. (\ref{47}) is the biconfluent Heun's differential equation and H(r) is the Heun polynomials.

The equation (\ref{46}) can be easily solve by using the Frobenius method as follow :
\begin{equation}
H (r) = \sum^{\infty}_{i=0} c_{i}\,r^{i}.
\label{48}
\end{equation}
Substituting Eq. (\ref{48}) into the Eq. (\ref{46}), we get the following recurrence relation for the coefficient:
\begin{equation}
c_{n+2}=\frac{1}{(n+2)(n+1+\gamma)}\,[\{q+\zeta\,(n+1)\}\,c_{n+1}-(\theta-2\,n)\,c_{n}].
\label{49}
\end{equation}
And the various coefficients are
\begin{equation}
c_1=\frac{q}{\gamma}\,c_0,\quad c_2=\frac{1}{2\,(1+\gamma)}\,[(q+\zeta)\}\,c_{1}-\theta\,c_{0}].
\label{50}
\end{equation}
The power series becomes a polynomial of degree $n$ by imposing the following two conditions:
\begin{equation}
c_{n+1}=0,\quad (\theta-2\,n)=0,\quad n=1,2...
\label{51}
\end{equation}
Using the above energy quantization condition we get the following eigenvalues equation : 
\begin{equation}
B^2+\frac{\lambda_s}{\beta}-4\,A\,D-2\,D=2\,n.
\label{52}
\end{equation}
Substituting $A, B, D$ into the above eigenvalue equation, we get the following eigenvalues of energy
\begin{eqnarray}
E_{n}&=&\frac{1}{2}\,[\alpha_0\,(2\,n+3+s)\pm \sqrt{\alpha^2_{0}\,(2\,n+3+s)^2+4\,k^2}]\nonumber\\
&=&\Omega\,(2\,n+3+s)\pm \sqrt{\Omega^2\,(2\,n+3+s)^2+k^2}.
\label{53}
\end{eqnarray}
The corresponding wave functions are  
\begin{equation}
\psi_{i\,n} (r)=r\,e^{-\frac{|l|}{\sqrt{2\,\Omega\,E_{n}}}\,r}\,e^{-\frac{r^2}{2}}\,H (r),
\label{54}
\end{equation}
where $l=\pm\,1,\pm\,2....\in {\bf Z}$ is integer.

If one set the constant $k=0$, we get the following energy levels from (\ref{53})
\begin{equation}
    E_{n}=2\,\Omega\,(2\,n+3+s)
\label{55}
\end{equation}
which is similar to the energy eigenvalues obtained in Ref. \cite{EPJC3} (see Eq. (46) in Ref. \cite{EPJC3} with $\xi=0$ there).

Now we evaluate the individual energy levels and wave-functions one by one by imposing the additional recurrence condition $c_{n+1}=0$ and others are in the same way. For example $n=1$ we have $c_2=0$ which implies from Eq. (\ref{50})
\begin{equation}
    E_{1}=\sqrt{2}\,\frac{l}{\alpha_0}
\end{equation}
the ground state energy levels. The corresponding ground state wave-function is given by
\begin{equation}
\psi_{i\,1} (r)=r\,e^{-\frac{|l|}{\sqrt{2\,\Omega\,E_{1}}}\,r}\,e^{-\frac{r^2}{2}}\,H (r),
\end{equation}

\section{Conclusions}

In this work, we have studied the Weyl fermions in a G\"{o}del-type space-time backgrounds. We have solved the Weyl equation in a linear class of G\"{o}del-type space-time and obtained the relativistic energy eigenvalues (\ref{53}) and wave-function (\ref{54}).

The energy levels obtained for the class of space-times has properties same to the the energy levels obtained in the case of quantum dynamics of Dirac fermions in the same space-time geometries in the theory of relativity \cite{EPJC3}. Note that the rotation of G\"{o}del space-time introduce a contribution in the Weyl energy levels. We claim the analogy between the energy levels for Weyl fermions in a linear class of G\"{o}del-type metrics and the Landau levels in curved spaces can be used to investigate the Weyl semimetals \cite{BY}. The systems investigated here can be used to describe condensed matter systems in curved geometries on the influence of rotation described by massless fermions \cite{Caval,GQ,JRFL,JRFL2,MMC}.


\begin{thebibliography}{99}

\bibitem{XCZ} X. C. Zhang, Q. W. Liu, C. S. Jia and L. Z. Wang, Phys. Lett. A {\bf 340}, 59 (2005).

\bibitem{LZY} L. Z. Yi, Y. F. Diao, J. Y. Liu and C. S. Jia, Phys. Lett. A {\bf 333}, 212 (2004).

\bibitem{AWT} A. W. Thomas, W. Weise, Structure of the Nucleon, Wiley, Berlin (2001).

\bibitem{BT} B. Thaller, The Dirac Equation, Springer, New York (1992).

\bibitem{TYW} T. Y. Wu and W. Y. P. Hwang, {\tt Relativistic quantum mechanics and quantum fields}, World Scientific, Singapore (1991).

\bibitem{WG} W. Greiner, {\tt Relativistic Quantum Mechanics: Wave Equations}, Springer, Berlin (2000).

\bibitem{JS} J. Santos, M. J. Rebouças, T. B. R. F. Oliveira and A. F. F. Teixeira, Eur.Phys.J. C {\bf 78}, 567 (2018), arXiv:1611.03985 [gr-qc].

\bibitem{CF} C. Furtado, T. Mariz, J. R. Nascimento, A. Yu. Petrov and A. F. Santos, Phys. Rev. {\bf D 79}, 124039 (2009), arXiv:0906.0554 [hep-th].

\bibitem{CF2} I. A. Pedrosa, A. Rosas and C Furtado, Int. J. Mod. Phys. : Conf. Ser. {\bf 18}, 140 (2012).

\bibitem{CF3} C. Furtado, J. R. Nascimento, A. Yu. Petrov and A. F. Santos, Phys. Rev. {\bf D 84}, 047702 (2011), arXiv:1106.4003 [hep-th].

\bibitem{JB3} J. B. Fonseca-Neto, A. Yu. Petrov and M. J. Reboucas, Phys. Lett. {\bf B 725}, 412 (2013), 	arXiv:1304.4675 [astro-ph.CO].

\bibitem{JA} J. A. Agudelo, J. R. Nascimento, A. Yu. Petrov, P. J. Porfírio and A. F. Santos, Phys. Lett. {\bf B 762}, 96 (2016), arXiv: 1603.07582 [hep-th].

\bibitem{Caval} E. Cavalcante, J. Carvalho and C. Furtado, Eur. Phys. J. Plus {\bf 131} , 288 (2016), 	arXiv:1603.04956 [quant-ph].

\bibitem{GC} F. Ahmed, Grav. Cosmol. {\bf 26} (2), 136 (2020). 

\bibitem{GC2} F. Ahmed, Grav. Cosmol. {\bf 26} (3), 265 (2020).

\bibitem{AHEP} T. P. Kling, F. Ahmed and M. Lalumiere, Adv. High. Energy Phys {\bf 2020}, 8713756 (2020), arXiv:2005.03417 [gr-qc].

\bibitem{BDB} B. D. B. Figueiredo, I. D. Soares, J. Tiomno, Class. Quantum Grav. {\bf 9}, 1593 (1992).

\bibitem{ND} N. Drukker, B. Fiol and J. Simon, JCAP (2004) {\bf 0410} : 012, arXiv:hep-th/0309199.

\bibitem{ND2} N. Drukker, B. Fiol and J. Simon, Phys. Rev. Lett. {\bf 91}, 231601 (2003), arXiv:hep-th/0306057.

\bibitem{SD} S. Das and J. Gegenberg, Gen. Relativ. Grav. {\bf 40}, 2115 (2008), arXiv:hep-th/0407053.

\bibitem{Wang} Z. Wang, Z.-W. Long, C- Y. Long and M. -L. Wu, Eur. Phys. J. Plus {\bf 130}, 36 (2015).

\bibitem{JC} J. Carvalho, A. M. de M. Carvalho and C. Furtado, Eur. Phys. J. C {\bf 74}, 2935 (2014), 	arXiv:1401.4217 [hep-th] .

\bibitem{VM} V. M. Villalba, Mod. Phys. Lett. A {\bf 08}, 3011 (1993).

\bibitem{SGF} S. G. Fernandes, G. de A. Marques and V. B. Bezerra, Class. Quantum Grav. {\bf 23}, 7063 (2006).

\bibitem{AH} A. Havare and T. Yetkin, Class. Quantum Grav. {\bf 19}, 1 (2002), arXiv:gr-qc/0108050v2.

\bibitem{GQ} G. Q. Garcia, J. R. de Oliveira, K. Bakke and C. Furtado, Eur. Phys. J. Plus {\bf 132}, 123 (2017), arXiv:1607.00456 [hep-th].

\bibitem{GQG} G. Q. Garcia, J. R. de S. Oliveira and C. Furtado, Int. J. Mod. Phys. D {\bf 27}, 1850027 (2017), arXiv:1705.10631 [hep-th]. 

\bibitem{EPJC} F. Ahmed, Eur. Phys. J. C {\bf 78}, 598 (2018).

\bibitem{EPJC2} F. Ahmed, Eur. Phys. J. C {\bf 79}, 104 (2019).

\bibitem{EPJC3} F. Ahmed, Eur. Phys. J. C {\bf 79}, 534 (2019).

\bibitem{MPLA} F. Ahmed and H. Hassanabadi, Mod. Phys. Lett. {\bf A 35}, 2050031 (2020), arXiv:1903.09311 [gr-qc].

\bibitem{CTP} F. Ahmed, Commun. Theor. Phys. {\bf 72}, 025103 (2020), arXiv:2002.03943 [hep-th]. 

\bibitem{Y} Y. Yang, Z. -W. Long, Q.-K. Ran, H. Chen, Z. -L. Zhao and C. -Y. Long, Int. J. Mod. Phys. A {\bf 36} (03), 2150023 (2021).

\bibitem{BI} B. I. Halperin, Phys. Rev. B {\bf 25}, 2185 (1982).

\bibitem{Naka} M. Nakahara, {\tt Geometry, Topology and Physics}, Cambridge University Press, Cambridge, (1982).

\bibitem{Weinberg} S. Weinberg, {\tt Gravitation and Cosmology: Principles and Applications of the General Theory of Relativity}, Wiley, New York (1972).

\bibitem{Cartan} E. Cartan, {\tt The Theory of Spinors}, Dover Publications, New York (1981).

\bibitem{Faiz} F. Ahmed, Commun. Theor. Phys. {\bf 68}, 735 (2017), arXiv: 1712.01274 [gr-qc].

\bibitem{Pal} P. Pal, Amer. Jour. Phys. {\bf 79}, 485 (2011), arXiv:1006.1718 [hep-ph].

\bibitem{BY} B. Yan and C. Felser, Ann. Rev. Condens. Matter Phys. {\bf 8}, 337 (2017), arXiv:1611.04182 [cond-mat.mtrl-sci].

\bibitem{JRFL} J. R. F. Lima, J. Brandao, M. M. Cunha and F. Moraes, Eur. Phys. J. D {\bf 68}, 94 (2014), 	arXiv:1405.6633 [cond-mat.mes-hall].

\bibitem{JRFL2} J. R. F. Lima and F. Moraes, Solid State Commun. {\bf 201}, 82 (2015), arXiv:1411.2826 [cond-mat.mtrl-sci].

\bibitem{MMC} M. M. Cunha, J. Brand˜ao, J. R. F. Lima and F. Moraes, Eur. Phys. J. B {\bf 88}, 288 (2015).

\end{thebibliography}
\end{document}